\documentclass[sn-mathphys,Numbered]{sn-jnl}

\usepackage{graphicx}%
\usepackage{multirow}%
\usepackage{amsmath,amssymb,amsfonts}%
\usepackage{amsthm}%
\usepackage{mathrsfs}%
\usepackage[title]{appendix}%
\usepackage{xcolor}%
\usepackage{textcomp}%
\usepackage{manyfoot}%
\usepackage{booktabs}%
\usepackage{algorithm}%
\usepackage{algorithmicx}%
\usepackage{algpseudocode}%
\usepackage{listings}%
\usepackage{siunitx}

\unnumbered

\begin{document}

\title[High Frequency Magnetometry with an Ensemble of Spin Qubits in Hexagonal Boron Nitride]{High Frequency Magnetometry with an Ensemble of Spin Qubits in Hexagonal Boron Nitride}


\author[1]{\fnm{Charlie J.} \sur{Patrickson}}

\author[1]{\fnm{Simon} \sur{Baber}}

\author[1]{\fnm{Blanka B.} \sur{Ga\'al}}

\author[2]{\fnm{Andrew J.} \sur{Ramsay}}

\author*[1]{\fnm{Isaac J.} \sur{Luxmoore}}\email{i.j.luxmoore@exeter.ac.uk}

\affil[1]{\orgdiv{Department of Engineering}, \orgname{University of Exeter}, \orgaddress{\postcode{EX4 4QF}, \country{UK}}} 

\affil[2]{\orgdiv{Hitachi Cambridge Laboratory}, \orgname{Hitachi Europe Ltd.}, \orgaddress{\postcode{CB3 0HE}, \country{UK}}} 


\abstract{Sensors based on spin qubits in 2D crystals offer the prospect of nanoscale sensing volumes, where the close proximity of the sensor and source could provide access to otherwise inaccessible signals. For AC magnetometry, the sensitivity and frequency range is typically limited by the noise spectrum, which determines the qubit coherence time. This poses a problem for III-V materials, as the non-zero spin of the host nuclei introduces a considerable source of magnetic noise. Here, we overcome this with a sensing protocol based on phase modulated continuous concatenated dynamic decoupling, which extends the coherence time towards the $T_1$ limit at room temperature and enables tuneable narrowband AC magnetometry. We demonstrate the protocol with an ensemble of negatively charged boron vacancies in hexagonal boron nitride, detecting in-plane AC fields within $\pm 150~\mathrm{MHz}$ of the electron spin resonance, and out-of-plane fields in the range of $\sim10-150~\mathrm{MHz}$. We measure an AC magnetic field sensitivity of $\sim1~\mathrm{\mu T/\sqrt{Hz}}$ at $\sim2.5~\mathrm{GHz}$, for a sensor volume of $\sim0.1~\mathrm{\mu m^3}$, and demonstrate that the sensor can reconstruct the AC magnetic field from a wire loop antenna. This work establishes the viability of spin defects in 2D materials for high frequency magnetometry, demonstrating sensitivities that are comparable to nitrogen vacancy centres in diamond for microscopic sensing volumes, and with wide-ranging applications across science and technology.}

\maketitle

\section*{Introduction}\label{sec:level1}

Solid-state spin systems are a promising platform for the development of novel magnetic field sensors, with applications ranging from the the pursuit of quantum technologies to reaction monitoring in cells \cite{Arai2022, Hatano2022, Cao2020, Shao2016, Schirhagl2014}. Particular progress has been made with the nitrogen vacancy in diamond, thanks to long spin coherence times under ambient conditions. This has enabled significant progress, with notable highlights such as, single NV scanning magnetometers with nanoscale resolution\cite{Rondin_RepProgPhys_2014,Balasubramanian_Nature_2008,Maze2008, Huxter2023, Gross2017}, sub-millihertz resolution magnetic resonance spectroscopy\cite{Schmitt_Science_2017,Boss_Science_2017,Glenn_Nature_2018} and ensemble based devices with sensitivities in the $~\mathrm{pT/\sqrt{Hz}}$ range for signal frequencies ranging from DC\cite{Fescenko_PhysRevResearch_2020} and low frequency \cite{Wolf_PhysRevX_2015,Eisenach_NatComms_2021} to GHz\cite{,Wang2022,alsid_arXiv_2022}.

As the sensitivity depends on the spin coherence time, dynamical decoupling is often employed to shield the qubit from sources of noise, extending coherence times towards the $T_1$ limit. Pulsed variations have achieved sensitivities in the $\mathrm{nT/\sqrt{Hz}}$ range\cite{Farfurnik2018, Pham2012}, however these are susceptible to errors arising from imperfect pulses \cite{Ishikawa2018, Cao2020}. Continuous dynamical decoupling \cite{Loretz2013, Kong2018, FortschrPhys2017_Cohen, Wang2021, Farfurnik2017, Cai2012}, on the other hand, avoids these issues and can also benefit from reduced power requirements\cite{Cao2020}.

Recently explored spin defects in hexagonal Boron Nitride offer an interesting alternative. To date, the most well-studied spin system in hBN is the negatively charged boron vacancy\cite{Gottscholl2020} ($V_B^-$), which has been investigated as a sensor of temperature\cite{Gottscholl2021}, strain\cite{Lyu_NanoLett_2022,Yang2022} and magnetic field\cite{Healey_NatPhys_2023,Huang_NatComms_2022,Kumar_PhysRevApplied_2022,Gottscholl2021,Rizzato2022}. A potential limitation is the relatively short spin echo time\cite{Ramsay2023, Haykal2021}. However, this can be mitigated with dynamic decoupling schemes\cite{Ramsay2023,Gong2022,Rizzato2022} and the 2D nature of the host material provides a unique opportunity for the sensor to be in close proximity to, or even embedded within, the system of interest\cite{Healey_NatPhys_2023,Huang_NatComms_2022,Kumar_PhysRevApplied_2022}.

In this work, we implement a continuous concatenated dynamic decoupling (CCDD) scheme \cite{FortschrPhys2017_Cohen} with an ensemble of boron vacancies, to simultaneously provide robust protection against dephasing, whilst also enabling high frequency magnetometry. This approach uses the CCDD drive field to manipulate spin state transitions into resonance with a signal field. We characterise the sensors performance, demonstrating the detection of magnetic fields in the tens of MHz to several GHz range. For a fixed static magnetic field, the sensor has a bandwidth $>200~\mathrm{MHz}$ and a sensitivity of $\sim1~\mathrm{\mu T/\sqrt{Hz}}$. We use the sensor to spatially map the magnetic field from a loop antenna and find good agreement between measurement, simulation and analytical solutions.

\section{Theoretical Background}
The AC magnetic field sensor uses an ensemble of spin-1 negatively charged Boron vacancies in an hBN flake. The spin has a highly nonlinear response to an ac-magnetic field. This is used to mix the frequency of a signal field with the electron spin resonance, and an additional drive field. Optimising the drive field then produces a DC component of the signal field that can be detected as a change in fluorescence, known as optically detected magnetic resonance, thereby sensing the signal. The benefit of using this scheme for AC magnetometry is twofold. Firstly, the stabilising drive fields improve coherence times, increasing the sensitivity. Secondly, the resonant frequencies of the sensor can be tuned using a static magnetic field and the parameters of the drive field. As we shall see in the following sections, this results in a narrow sensor bandwidth, capable of determining the amplitude and frequency of an unknown signal in the MHz to GHz range.


\subsection{\label{sec:level2}{Structure of the Boron Vacancy}}
\begin{figure*} 
\includegraphics[width=1\columnwidth]{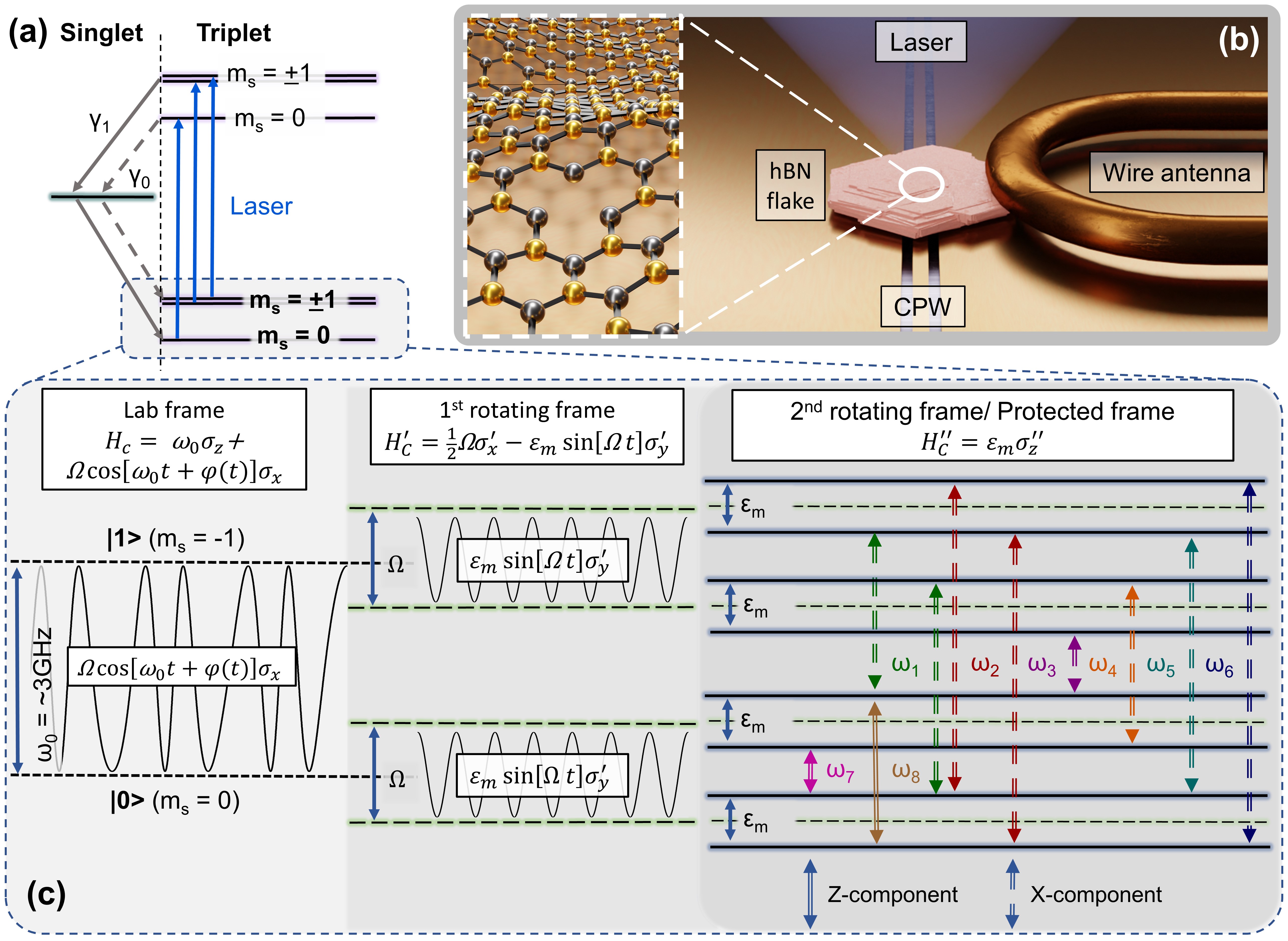}
\caption{\label{fig1} \textbf{(a)} Simplified energy level diagram of the $V_B^-$ defect in hexagonal boron nitride. \textbf{(b)} Illustration of the sensor and experimental setup with an inset depicting the Boron vacancy. \textbf{(c)} Energy levels of the triplet ground state spin system under phase modulated concatenated dynamic decoupling. The concept of CCDD can be understood by viewing the system in frames of reference that rotate at the same frequency as the resonant MW drive fields; the first frame rotates at $\omega_0\sigma_z$, the second rotates at $\Omega\sigma_x^{\prime}$. Here we find static fields proportional to the amplitudes of the drive terms; $\Omega$ in the first frame, $\epsilon_m$ in the second. The sensing protocol relies on selection of the drive fields such that the signal field is resonant with one of the eight resonances $\omega_1$ - $\omega_8$, thereby driving that transition, resulting in a detectable change in spin state (see equations \ref{eq:ProtectedPerpFrameSignal} and \ref{eq:ProtectedParallelFrameSignal}).}
\end{figure*}

The energy level diagram of the negatively charged boron vacancy is shown in Fig. \ref{fig1}(a). Two unpaired electrons form a radiative spin triplet system (total spin quantum number, S = 1), with optical ground state zero-field splitting (ZFS) of $D_{GS} \approx +3.5~\mathrm{GHz}$ \cite{Gottscholl2020}. The intersystem crossing rate, from triplet to singlet, is spin dependent ($\gamma_1 \approx 2\gamma_0$) \cite{NanoLett2022_Baber}, which enables initialisation of the ground state spin in $m_s=0$ via optical pumping, and spin readout via the photoluminescence (PL) intensity.

The electron Hamiltonian, $H_e$ can be described by;
\begin{eqnarray}
H_e= DS^2_z + E(S^2_x - S^2_y) + \gamma_eB_zS_z + H_c(t) + \delta H(t)
\label{eq:electron Hamiltonian}
\end{eqnarray}
where $D=3.45~\mathrm{GHz}$ and $E=59~\mathrm{MHz}$ describe the ZFS terms in our sample\cite{Ramsay2023}, $\gamma_e \approx28~\mathrm{MHz/mT}$ is the electron gyromagnetic ratio and $S_i$ are the electron spin-1 operators. We apply a static magnetic field, $B_z$, parallel to the hBN c-axis. This separates the $\vert m_s=0\rangle \leftrightarrow \vert m_s=\pm 1\rangle$ transitions. For a narrowband external ac magnetic field $H_c(t)$  near resonance with the energy gap $\omega_0$ between the ground  $\vert m_s=0\rangle$ and $\vert m_s=-1\rangle$ states, the $\vert m_s=+1\rangle $ state can be neglected, reducing the system to a spin-half or qubit, with Hamiltonian $H_e\rightarrow \frac{\omega_0}{2}\sigma_z$. The final term, $\delta H(t)$ describes low frequency magnetic noise dominated by the electron-nuclear interaction. This limits the spin-echo times to under 100 ns \cite{Haykal2021,Ramsay2023}, and therefore dictates the sensitivity to external magnetic fields.   

\subsection{\label{sec:level3}Phase-modulated Dynamic Decoupling}
In our previous work, we showed that decoherence arising from magnetic noise can be mitigated in $V_B^-$ ensembles with CCDD\cite{Ramsay2023}. Here, we  choose phase, rather than amplitude, modulated CCDD as it requires less power and has shown superior performance in extending the coherence time\cite{Ramsay2023}. The scheme is illustrated in Fig. \ref{fig1}(c), where a linearly polarised MW field of amplitude $\Omega$ is applied along the $x$ axis to drive a Rabi oscillation between $\vert 0\rangle$ and $\vert 1\rangle$. A second drive is applied as an additional phase term, $\phi(t) = \frac{2\epsilon_m}{\omega_m}\sin{(\omega_m t-\theta_m)}$. With this, we arrive at a MW control Hamiltonian of the form \cite{Wang2020, FortschrPhys2017_Cohen, Farfurnik2017}:

\begin{eqnarray}
H_c= \frac{1}{2}\omega_0\sigma_z\ + \Omega\cos{(\omega t - \phi(t)})\sigma_x\ \ \ 
\label{eq:lab frame}
\end{eqnarray}

Insight can be gained by viewing the system in a frame of reference rotating at the drive frequency $\omega$.  Here, we find $H_c'=e^{i\omega t\sigma_z/2}H_ce^{-i\omega t\sigma_z/2} - \omega\sigma'_z/2$

\begin{eqnarray}
H_c'= \frac{1}{2}(\omega_0-\omega)\sigma_z^\prime+\frac{1}{2}\Omega\sigma_x^\prime + \frac{1}{2}\Omega\phi(t)\sigma_y^\prime \ \ 
\label{eq:1stRotatingFrame}
\end{eqnarray}

where $^\prime$ refers to axes in the first rotating frame, we have applied the RWA and take the limit $\epsilon_m\ll\Omega$. In the case of a single drive ($\epsilon_m = 0$), resonant with the electron spin resonance ($\omega=\omega_0$), we find a static energy gap of $\Omega\sigma_x^\prime$ – the eigenvalues of the first dressed state. By selecting $\omega_m=\Omega$, in the case of a double drive ($\epsilon_m \ne 0$) the phase modulation term, $\phi(t)$, drives transitions between the first dressed states, shown in the middle panel of Fig. \ref{fig1}(c). 

Moving to a second frame of reference rotating around the $x^{\prime}$-axis at the $\phi(t)$ frequency $\omega_m$, we find

\begin{eqnarray}
H_c''= \frac{1}{2}(\Omega - \omega_m)\sigma_x^{\prime\prime} + \frac{1}{2}\epsilon_m(\sigma_y^{\prime\prime}\sin(\theta_m) + \sigma_z^{\prime\prime}\cos(\theta_m))\ \ \ 
\label{eq:ProtectedFrame}
\end{eqnarray}

where we have again applied the RWA, taken $\phi=0$ for simplicity and the limit $\epsilon_m\ll\Omega$. We refer to this second rotating frame of reference, shown in the right hand panel of Fig. \ref{fig1}(c), as the \textit{protected frame}. This space is robust against the dominant sources of noise, as demonstrated in the following section. Setting $\omega_m = \Omega$ results in a static energy gap of amplitude $\epsilon_m$ in the protected frame, with an axis that can be selected using the phase of the second drive term, $\theta_m$. Here, an ideal Rabi oscillation is represented by the spin-vector $\sigma^{\prime\prime}=(0,0,-1)$. 



Overall, this frame describes a pair of spin states, separated by an energy gap of $\epsilon_m$ that rotate around both the $z$ and $x^{\prime}$ axes, at frequencies $\omega$, and $\omega_m$. This results in states that are dynamically decoupled from both bit and phase-flip errors, extending the coherence times towards the $T_1$ limit. 

\subsection{\label{sec:level4}AC Magnetometry}
In the lab frame, an AC signal of frequency $\omega_s$ is described by an additional drive term, $H_{AC}$, in the Hamiltonian.

\begin{eqnarray}
H_{AC\perp}= g_{\perp} \sigma_x \cos(\omega_st + \phi_s)\\
H_{AC\parallel} = g_{\parallel} \sigma_z \cos(\omega_st + \phi_s)
\label{eq:LabFrameSignal}
\end{eqnarray}

where $g_{\parallel} = \frac{1}{2}\gamma_eB_\parallel$ and $g_{\perp} = \frac{1}{2}\gamma_eB_\perp$. Transforming to the protected frame, the signal term becomes: 
\begin{eqnarray}
H''_{AC\perp}= \frac{1}{4}g_\perp[\sigma_z^{\prime\prime}\cos((\omega_s - \omega)t + \phi_s) - \sum_{\gamma = +,-}S_{\gamma}e^{\gamma i((\omega_s - \omega - \gamma\omega_m)t + \phi_s)} + h.c.],
\label{eq:ProtectedPerpFrameSignal}
\end{eqnarray}

\begin{eqnarray}
H''_{AC\parallel}= \frac{1}{2}g_\parallel[ \sum_{\gamma = +,-}S_{\gamma}e^{-i((\omega_s - \gamma \omega_m)t + \phi_s)} + h.c.].
\label{eq:ProtectedParallelFrameSignal}
\end{eqnarray}

In the protected frame, the signal term can drive a Rabi oscillation when one of the shifted frequencies is resonant with the $\epsilon_m$ splitting of the protected spin, and can be detected via spin dependent PL. This yields eight resonances at frequencies of $\epsilon_m$: $\omega_s=\omega_0\pm\epsilon_m$, $\omega_s=\omega_0\pm\epsilon_m+\omega_m$ and $\omega_s=\omega_0\pm\epsilon_m-\omega_m$ for in-plane signals, and $\omega_s=\omega_m\pm\epsilon_m$ for out of plane signals (see $\omega_1$ to $\omega_8$ in Fig. \ref{fig1}(c)). 
The signal frequency can be determined by tuning the resonances via the CCDD drive parameters. Moreover, because the resonance conditions hold for any value of the signal phase $\phi_s$, phase locking between the signal and microwave control waveform is not necessary.

The magnetic noise can be described as $\delta H(t) = \sum_{\omega_{\gamma},\alpha}\sigma_{\alpha}B_{N,\alpha}(\omega_{\gamma})\cos(\omega_{\gamma} t)$. This takes the same form as $H_{AC}$ and transforms to the protected frame in the same way.  By optimising the CCDD scheme, the noise spectrum can be frequency shifted away from resonance with the qubit, reducing decoherence. The primary source of noise is low frequency electron-nuclear interactions along the z-axis.  In the protected frame, decoupling from nuclear noise requires $\omega_\gamma \pm \omega_m \neq \epsilon_m$, which holds for low frequencies of $\omega_{\gamma}$ associated with electron-nuclear coupling. The second largest source of noise comes from slow fluctuations in drive amplitude, $\Omega$, where the requirements are $\omega_\gamma -\omega_0 \pm \omega_m, \omega_\gamma +\omega_0 \pm \omega_m, \omega_{\gamma} \pm\omega_0 \neq \epsilon_m$.

\section{Experimental Implementation}
\subsection{\label{sec:level5}Sample and Experimental Setup}
In our experiments we use an ensemble of Boron vacancies in an hBN flake, transferred to a co-planar waveguide on a sapphire substrate\cite{NanoLett2022_Baber,Ramsay2023} (see Fig. \ref{fig1}(a)). The MW control fields are applied to the CPW using an arbitrary waveform generator (AWG). To demonstrate that the scheme does not require phase locking between signal and control fields, we use a separate signal generator to drive the signal, which we apply to either the CPW or, in later experiments, to a $40~\mathrm{\mu m}$ diameter wire mounted on an XYZ translation stage (Fig. \ref{fig1}(b)). PL ($>750~\mathrm{nm}$) is excited using a 488 nm laser and detected with a single photon avalanche diode (see methods for further experimental details).

\subsection{\label{sec:level6}AC Field Sensing Protocol}
\begin{figure} 
\includegraphics[width=0.5\textwidth]{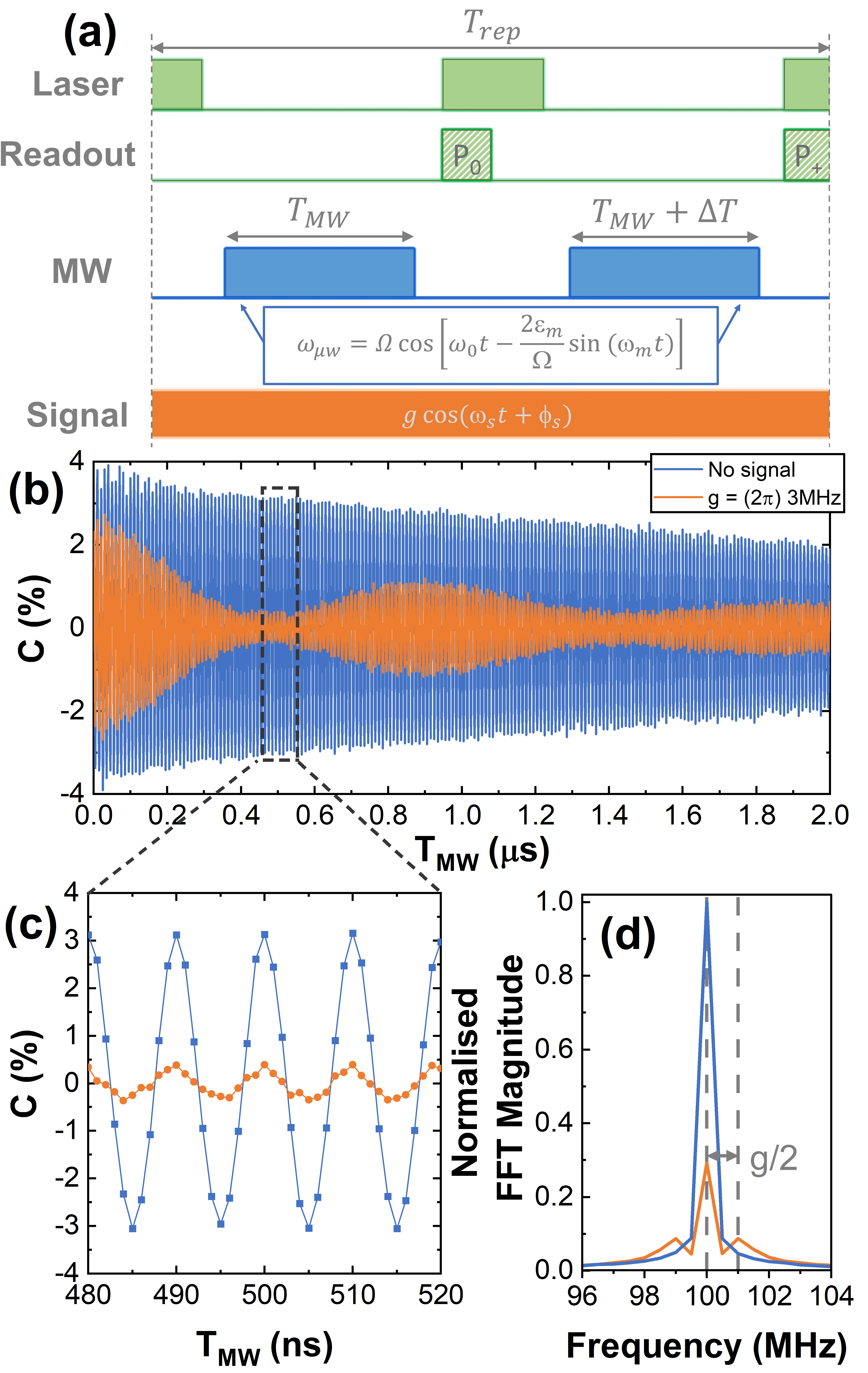}
\centering
\caption{\label{fig2} AC Field Sensing Protocol using signal driven Rabi-oscillations of the CCDD protected qubit. \textbf{(a)} Optimised pulse-sequence for AC field sensing. Alternating microwave pulses of length $T_{MW}$ and $T_{MW}+\Delta T$ are interlaced with laser pulses for preparation and readout of the spin. We define contrast as $(P_{0}-P_{+})/P_{+}$; $P_0$ ($P_+$) is the measured PL following a microwave pulse of length $T_{MW}$ ($T_{MW}+\Delta T$.). For AC sensing we choose $\Delta T = 1/(2\Omega)$. \textbf{(b)} Measurement using the pulse sequence described in \textbf{(a)}, highlighting the CCDD stabilised Rabi-oscillation ($\Omega =(2\pi)100~\mathrm{MHz}$ and $\epsilon_m=(2\pi)30~\mathrm{MHz}$). Shown with and without signal, $H_{AC\perp}$, with $\omega_s=(2\pi)2.577~\mathrm{GHz}$. \textbf{(c)} Close up of \textbf{(b)} around $T_{MW}=500~\mathrm{ns}$. \textbf{(d)} Fourier transforms of \textbf{(b)}. The signal driven Rabi oscillation in the protected frame manifests as the sidebands of a Mollow triplet.}
\end{figure}

The pulse sequence used for AC-field sensing is illustrated in Fig. \ref{fig2}(a) and consists of two repeating sequences of qubit initialisation, spin manipulation, and readout. The first laser pulse initialises the ensemble into the $\vert0\rangle$ state. Next, the spin is manipulated with a MW control sequence, before a second laser pulse excites PL, which is recorded as $P_0$. This laser pulse also reinitialises the spin for the second MW manipulation, which is again followed by a measurement of the PL, $P_+$. In the two sequences, the waveform $\omega_{MW}=\Omega \cos[\omega_0 t - \frac{2\epsilon_m}{\Omega} \sin(\omega_m t)]$ is applied with $\Omega=\omega_m=(2\pi)100~\mathrm{MHz}$, but for total times, $T=T_{MW}$ and $T=T_{MW} + \Delta T$, respectively. We use the contrast, $C = (P_0 - P_+)/ P_+$, as the sensor readout parameter. A typical measurement of the contrast is plotted in blue in \ref{fig2}(b), where $\Delta T=1/(2\Omega)=5~\mathrm{ns}$ is fixed and $T_{MW}$ is swept, effectively measuring the derivative of the CCDD stabilised Rabi-oscillation. This has the advantage of cancelling the effect of $T_1$ relaxation\cite{Ramsay2023} and maximizes the contrast, and therefore sensitivity to an external signal.

In the presence of an in-plane continuous external signal applied via the CPW with $\omega_s=\omega_0+\epsilon_m=(2\pi)2.557~\mathrm{GHz}$ the sensor undergoes two Rabi cycles at two different Rabi frequencies, simultaneously. The first is the CCDD stabilized lab frame Rabi oscillation with frequency $\Omega$. The second is driven by the signal field in the protected frame, with a Rabi frequency proportional to the signal amplitude $g_\perp$. The difference in Rabi frequencies produces a beating effect, as shown in orange in Fig. \ref{fig2}(b). In the Fourier domain, this manifests as a Mollow triplet centered on the lab frame Rabi frequency $\Omega$, with two sidebands offset by the Rabi frequency of the signal in the protected frame, $g/ 2$ (Fig. \ref{fig2}(d)). This provides a method to determine the signal amplitude, which is used to calibrate the sensor (see Supplementary Information section \ref{sec:13}).

\subsection{\label{sec:7}{Sensor Performance}}

\begin{figure*}
\includegraphics[width=1\columnwidth]{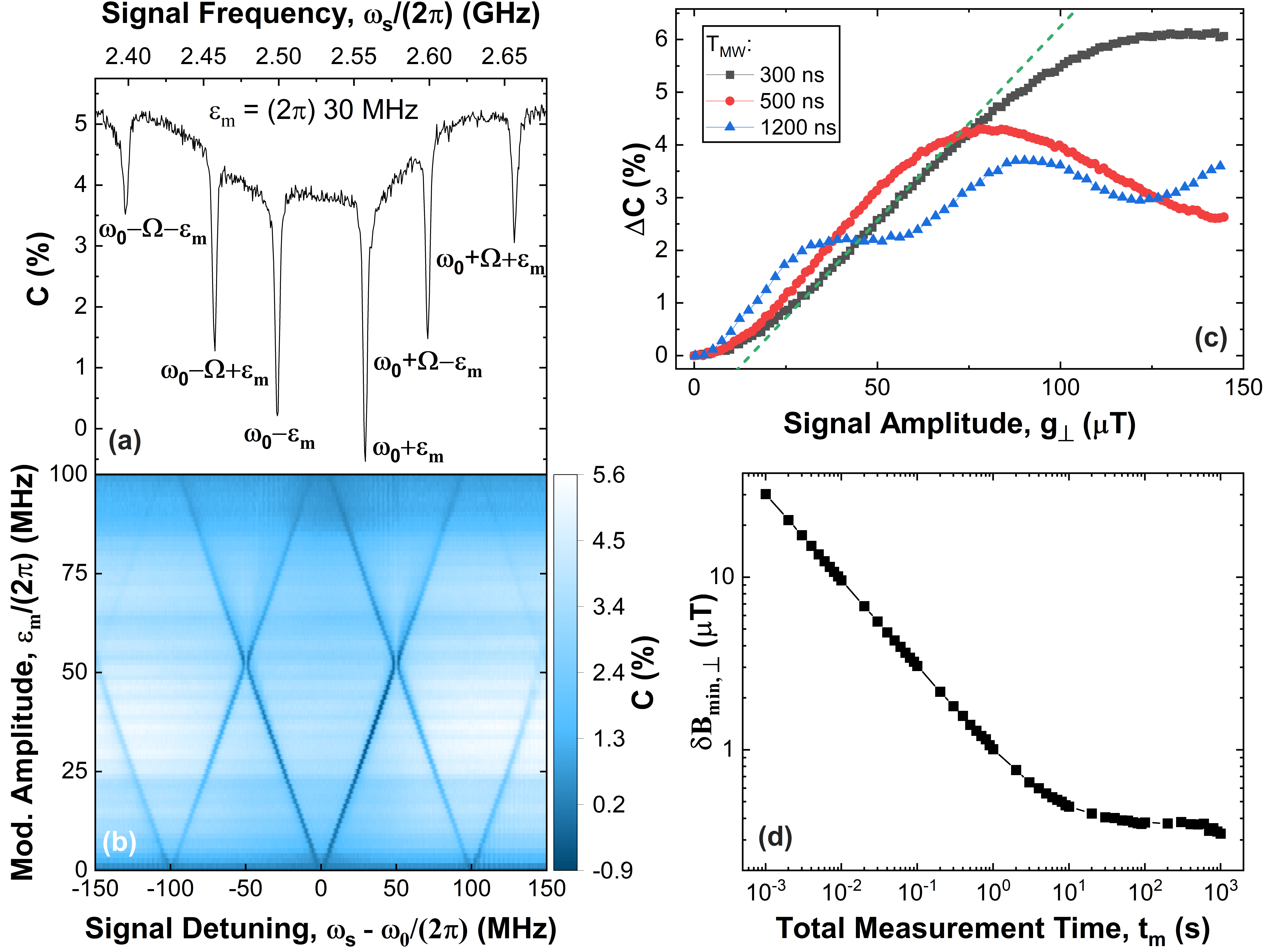}
\caption{\label{fig3} Sensor response to in-plane AC magnetic fields, $g_\perp$. \textbf{(a)} Sensor frequency response for Rabi Frequency, $\Omega=(2\pi)100~\mathrm{MHz}$ and modulation amplitude, $\epsilon_m=(2\pi)30~\mathrm{MHz}$, showing narrow bandwidth response at frequencies $\omega_s=\omega_0\pm\epsilon_m$ and $\omega_s=\omega_0\pm\Omega\pm\epsilon_m$. \textbf{(b)} Tuning of sensor frequency response with $\epsilon_m$. \textbf{(c)} Sensor amplitude response, in terms of the change in contrast, $\Delta C$, for different sensing times, $T_{MW}$. The signal frequency is tuned to $\omega_s=\omega_0+\epsilon_m=(2\pi)2.557~\mathrm{GHz}$. The dashed green line shows the maximum slope, $\max\lvert\frac{\partial (\Delta C)}{\partial B}\rvert$ for $T_{MW}=300~\mathrm{ns}$. \textbf{(d)} Minimum resolvable magnetic field, $\delta B_{min,\perp}$, as a function of measurement integration time.}
\end{figure*}

To evaluate the sensing protocol, in Fig. \ref{fig3} we quantify the sensor performance in response to an in plane signal, $g_\perp$, delivered through the CPW. Whilst the signal-driven Rabi-oscillation measurement in Fig. \ref{fig2}(b) can determine the signal amplitude and provides a means to calibrate the sensor, it is inconvenient to make such a measurement repeatedly, for example when determining the frequency or spatial extent of a signal. Instead, we employ a type of variance detection\cite{Degen2017}, where we specifically select $T_{MW}=N/\Omega$, where N is an integer, and $\Delta T = 1/2\Omega$ in order to compare adjacent peaks and troughs of the lab frame Rabi oscillation, thereby maximising the readout contrast, and hence the sensitivity.

We begin by measuring the contrast as a function of signal frequency, with fixed CCDD parameters $\Omega = \omega_m = (2\pi)100~\mathrm{MHz}$, $\epsilon_m = (2\pi)30~\mathrm{MHz}$, $T_{MW} = 300~\mathrm{ns}$ and $\Delta T = 5~\mathrm{ns}$ (Fig. \ref{fig3}(a)). Sweeping the signal frequency in a 300 MHz range around the electron spin resonance at $\omega_0$, the sensor detects six resonances, as predicted in Eq. \ref{eq:ProtectedPerpFrameSignal}. Using a Lorentzian fit the average linewidth is 3.9 MHz, which determines the frequency resolution of the sensor. In Fig. \ref{fig3}(b) we show how an unknown signal frequency can be determined. Whilst recording the contrast, the CCDD drive is used to tune the sensor into resonance with the signal by sweeping $\epsilon_m$. The six resonances shift as expected, and we achieve a detectable frequency range $\pm 150~\mathrm{MHz}$ with respect to $\omega_0$. In principle, a similar measurement can be made by sweeping $\Omega$ and the frequency range can also be extended by changing the static magnetic field that determines $\omega_0$. The lower frequency limit on a detectable signal is expected to be around 200 MHz, due to increased decoherence close to the ground state level anti-crossing \cite{NanoLett2022_Baber, Mathur2022, Onizhuk2021}. The upper limit is bound by the frequency limit of the microwave electronics and the strength of the static magnetic field. However, it is realistic to expect that the sensor could operate in a similar manner for signals up to tens of GHz. 

Next, with the signal tuned to the $\omega_s = \omega_0+\epsilon_m \approx(2\pi)2.557~\mathrm{GHz}$ resonance, we quantify the amplitude sensitivity to an in-plane AC B-field. In Fig. \ref{fig3}(c) the change in contrast, $\Delta C=\lvert C_0-C_s \rvert$, where $C_s$ ($C_0$) is contrast with (without) the signal applied, is plotted as a function of the signal amplitude, for three different $T_{MW}$ (see Supplementary Information section \ref{sec:13} for signal calibration). The sensor is most sensitive, where the gradient, $\frac{\partial (\Delta C)}{\partial B}$ is maximum, which is shown for $T_{MW}=300~\mathrm{ns}$ as the green dashed line in Fig \ref{fig3}(c). From this, we calculate the minimum resolvable change in magnetic field \cite{Stark2017}, 

\begin{eqnarray}
\delta B_{min} (t_{m}) = \frac{\sigma(t_{m})}{\max\lvert\frac{\partial (\Delta C)}{\partial B}\rvert}
\end{eqnarray}

where $\sigma (t_m)$ is the standard deviation of $\Delta C$ as a function of the measurement time $t_{m}$. To measure $\sigma(t_{m})$, $\Delta C$ is repeatedly sampled at 1 ms time intervals and the resulting values of $\delta B_{min,\perp} (t_{m})$ are plotted in Fig. \ref{fig3}(d). $\delta B_{min,\perp}$ follows the expected square root dependence for Shot noise limited detection, up to total measurement times of tens of seconds, when drifting laser power becomes significant. For $t_m<10~\mathrm{s}$, the sensitivity, $\eta_{\perp} =\delta B_{min,\perp} (t_{m}) \sqrt{t_m}$ is $\sim1~\mathrm{\mu T/ \sqrt{Hz}}$, which is currently limited by the data rate of our time-tagging electronics (see methods).

The measured sensitivity is smaller than reported figures for DC magnetic field sensing using ensembles of boron vacancies, where values of $\sim85~\mathrm{\mu T/ \sqrt{Hz}}$ have been achieved \cite{Gottscholl2021}, but considerably larger than state-of-the-art NV-center based GHz frequency sensors, where sensitivities can reach the $\mathrm{pT/ \sqrt{Hz}}$ range \cite{Wang2022, alsid_arXiv_2022}. This difference in performance is largely due to a difference in sensor volume, $V$, with the diamond devices having volumes of $V\sim0.04~\mathrm{mm^3}$ \cite{Wang2022} and $0.63~\mathrm{mm^3}$ \cite{alsid_arXiv_2022}, compared to $\sim0.1~\mathrm{\mu m^3}$ in our case. Some applications preclude the use of large sensing volumes, for example when measuring signals from individual or small numbers of spins \cite{Taylor2008}, in which case scaling the sensitivity by the volume provides a more representative metric of performance. Assuming a uniform defect density, the scaled sensitivity, $\widetilde{\eta} = \eta \sqrt{V}$, gives figures of $\widetilde{\eta}\sim2~\mathrm{pT~ Hz^{-1/2}~mm^{3/2}}$ for the diamond sensors compared to $\widetilde{\eta}\sim10~\mathrm{pT~Hz^{-1/2}~mm^{3/2}}$ in this case, implying that significant gains in sensitivity can be made by increasing the sensor volume (and/or defect density). However, there is an inherent trade-off between spatial resolution and the sensor volume, with the requirements driven by the particular application. 

Fig. \ref{fig3}(c) also illustrates how $T_{MW}$ can be selected according to the required dynamic range. For small amplitude signals the sensitivity is improved with a longer $T_{MW}$, whereas for larger signal amplitudes, the monotonic range of the sensor can be increased by selecting a shorter $T_{MW}$. This is a direct consequence of the fact that we are measuring a signal driven Rabi oscillation; the larger the signal field, the faster the protected frame Rabi oscillation and the earlier we can detect a change in contrast, thereby affording a shorter $T_{MW}$.

\subsection{External AC Magnetic Field}

\begin{figure*} 
\includegraphics[width=0.5\textwidth]{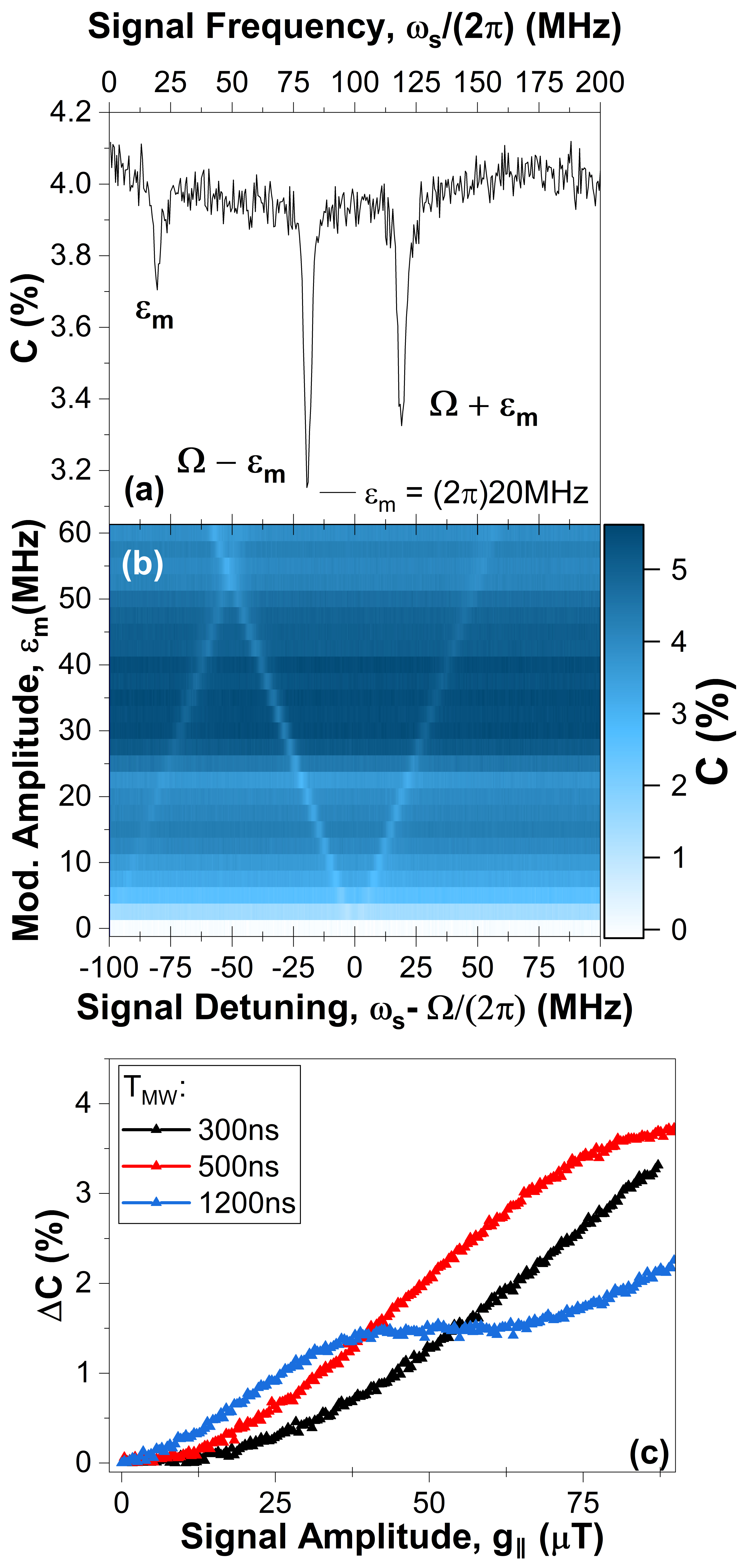}
\centering
\caption{\label{fig4} Sensor response to out of plane AC magnetic field, $g_\parallel$, from an external signal source. \textbf{(a)} Sensor frequency response for Rabi Frequency, $\Omega=(2\pi)100~\mathrm{MHz}$ and modulation amplitude, $\epsilon_m=(2\pi)20~\mathrm{MHz}$, showing narrow bandwidth response at $g_\parallel$-specific frequencies $\omega_s=\pm\epsilon_m$ and $\omega_s=\Omega\pm\epsilon_m$.\textbf{(b)} Tuning of sensor frequency response with $\epsilon_m$. \textbf{(c)} Sensor amplitude response for different $T_{MW}$ when $\omega_s=\Omega-\epsilon_m$.} 
\end{figure*}

Thus far, the experiments have been focused on signals applied from an independent source via the CPW, which is designed to achieve a large in-plane magnetic field. This simulates the case of a signal collected by an antenna and directed to the sensor. To simulate the direct detection of a signal, we instead use a nearby loop of $40\mathrm{\mu m}$ diameter wire mounted on an XYZ translation stage (Fig. \ref{fig1}(b)), driven by an independent signal generator. This is first used to characterise the sensor's response to out-of-plane fields, $g_\parallel$. We begin by positioning the wire adjacent to our device and for CCDD parameters of $\Omega = (2\pi)100~\mathrm{MHz}$, $\epsilon_m = (2\pi)30~\mathrm{MHz}$, $T_{MW} = 300~\mathrm{ns}$, we measure the contrast as a function of signal frequency (Fig. \ref{fig4}(a)). Three resonances are observed; $\omega_s=\Omega-\epsilon_m\approx(2\pi)71~\mathrm{MHz}$ and $\omega_s=\Omega+\epsilon_m\approx(2\pi)129~\mathrm{MHz}$ are predicted for $g_\parallel$ signal components in eq. \ref{eq:ProtectedParallelFrameSignal}, whereas $\omega_s=\epsilon_m\approx(2\pi)29~\mathrm{MHz}$ originates from a resonance in a third rotating frame with respect to $\omega_m \sigma_z^{\prime\prime}$ (see Supplementary Information section \ref{sec:12}). 

As with an in-plane signal, we are able to tune the frequency of these resonances with $\epsilon_m$, and therefore determine a signal's frequency. In Fig. \ref{fig4}(b), the contrast is plotted as a function of $\epsilon_m$ and $\omega_s$, revealing the expected linear dependence of the resonances. For out-of-plane magnetic fields, the upper limit of the frequency range is limited by $\Omega$, and therefore the available microwave power and the conversion efficiency of the CPW. The lower limit is currently determined by the CCDD protocol, which requires $\Omega>\epsilon_m$ and $\epsilon_m>\sim(2\pi)10~\mathrm{MHz}$ \cite{Ramsay2023}. This could be extended by concatenating a third drive term \cite{Cai_NJP_2012}, by including quantum frequency mixing\cite{Wang2022_PhysRevX} or continuous heterodyne detection\cite{Wang2022}. Fig. \ref{fig4}(c) plots the amplitude response of the sensor, which again shows similarities to in-plane field sensing (Fig. \ref{fig4}(c)), where the monotonic range and sensitivity (slope) are dependent on the $T_{MW}$ sensing time. 
\subsection{\label{sec:8}{Magnetic Field Mapping}}

\begin{figure*} 
\includegraphics[width=1\columnwidth]{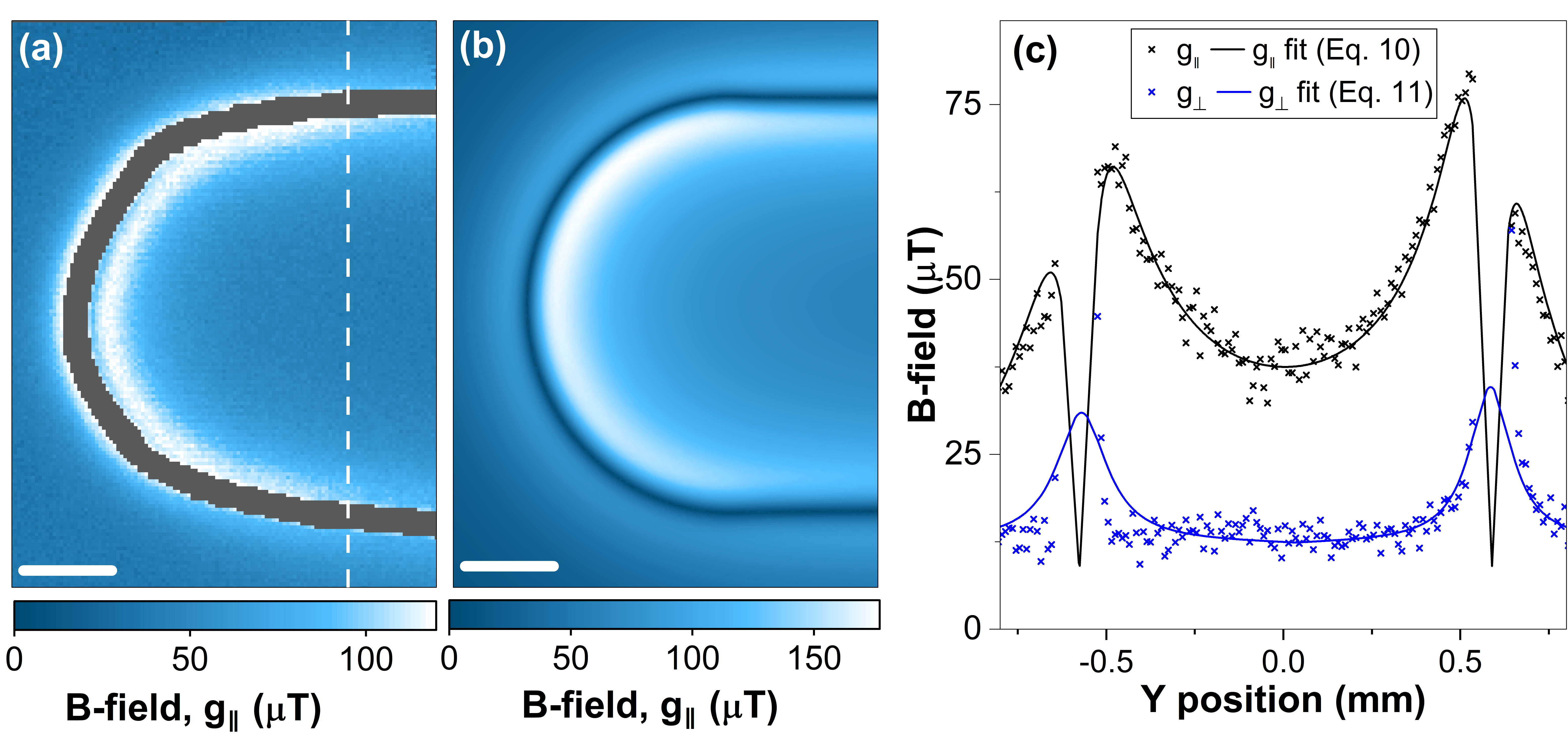}
\caption{\label{fig5} Magnetic field mapping of a resonant AC signal. \textbf{(a)} Measured and \textbf{(b)} simulated XY-maps of the out of plane magnetic field component, $g_\parallel$, from a $40\mathrm{\mu m}$ diameter wire loop. The wire was driven with a signal frequency $\omega_s = \Omega - \epsilon_m = (2\pi)71~\mathrm{MHz}$ to meet the resonance condition of the CCDD scheme. The sensor readout was integrated over $4~\mathrm{s}$ for each $10~\mathrm{\mu m}^2$ pixel in (a). The scale markers in (a) and (b) are $250~\mathrm{\mu m}$. Greyed-out areas correspond to sections where the wire obscures the photoluminescence collection path.  \textbf{(c)} Scanning across the two parallel sections of wire described by the linecut in (a), the in plane, $g_\perp$, and out of plane, $g_\parallel$, magnetic field components of the signal are mapped using the $g_\perp$ and $g_\parallel$-specific frequencies $\omega_s = \omega_0 + \epsilon_m = (2\pi)2580~\mathrm{MHz}$ and $\omega_s = \Omega - \epsilon_m = (2\pi)71~\mathrm{MHz}$, respectively. The solid lines show fits to the data using an analytical solution of the Biot-Savart law for two infinitely long parallel conductors.}
\end{figure*}

Finally, we provide a proof of principle demonstration of our hBN based sensor by mapping the magnetic field from the wire loop antenna. The sensor and readout laser spot are at fixed positions, whilst the antenna is positioned $\sim50~\mathrm{\mu m}$ above the sample and scanned in the XY-plane. For a resonant signal frequency, $\omega_s=\Omega-\epsilon_m=(2\pi)71~\mathrm{MHz}$ the magnetic field map is shown in Fig. \ref{fig5}(a). Missing data in grey corresponds to positions where the wire obscures the collection pathway. We compare this result to a Comsol simulation of the out-of-plane magnetic field, shown in Fig. \ref{fig5}(b), and find good qualitative agreement. In particular, at the inner-side of the apex of the loop, the signal amplitude can be seen to reach a maximum then dip before reaching the wire, which is specific to the out-of-plane magnetic field component.

Away from the apex, the wire loop approximates two parallel wires, which allows us to demonstrate the sensors capacity to differentiate between the in plane and out of plane components of the magnetic field. With the X-position fixed, as indicated by the dashed line in Fig. \ref{fig5}(a), the two parallel sections of wire are scanned along $x$, for signal frequencies $\omega_s = \Omega - \epsilon_m = (2\pi)71~\mathrm{MHz}$ and $\omega_s = \omega_0 + \epsilon_m = (2\pi)2.58~\mathrm{GHz}$, corresponding to $g_\perp$ and $g_\parallel$, respectively. The experimental results are shown in Fig. \ref{fig5}(c), alongside a fit to the analytical solution of the Biot-Savart law for two parallel conductors. We find good agreement to the experiment for sensor to source distances of $87\pm2\mathrm{\mu m}$ and $73\pm2\mathrm{\mu m}$ for each section of parallel wire. Note, the magnetic field amplitude for $g_\perp$ is small relative to $g_\parallel$. This is due to a reduced power output of the signal generator at higher frequencies, which is reflected in the current fit parameters $I_\perp = 16.2~\mathrm{mA}$ and $I_\parallel = 81.7~\mathrm{mA}$. The close agreement between experiment and model shows the hBN sensors capability to map the spatial variation of high frequency magnetic fields, which is relevant for a host of applications, for example in imaging microwave circuitry and probing the near-field of antennas \cite{Horsley2018}.

\section{Conclusions}
In this work, we have implemented a phase modulated CCDD sensing scheme with an ensemble of negatively charged boron vacancies in hBN, which is sensitive to both signal frequency and amplitude. The scheme avoids the timing errors associated with pulsed techniques, requires no phase matching of the signal field and benefits from sensitivities supported by coherence times approaching the $T_1$ limit, even at room temperature operation. In proof of principle experiments we have demonstrated a detectable frequency range of $\pm150~\mathrm{MHz}$ of the ESR for in plane fields and $\sim10 - 150~\mathrm{MHz}$ for out of plane fields. We achieve a sensitivity of $\sim1~\mathrm{\mu T/\sqrt{Hz}}$ at $\sim2.5~\mathrm{GHz}$ for in-plane fields, for an estimated sensor volume of only $\sim0.1~\mathrm{\mu m^3}$. The large Rabi-frequency, enabled by directly integrating the hBN layer with the CPW, means that the sensor can be tuned over a range $>250~\mathrm{MHz}$ for a fixed DC magnetic field. 


In future work, the sensitivity could be enhanced by increasing the intensity of the read-out signal. For example, by improving the excitation and collection efficiencies, increasing the sensor volume and/or defect density, or by using brighter defect species \cite{Guo2023, Stern2022}. The spatial resolution of the mapping setup used for Fig. \ref{fig5}, which is currently limited to $\sim0.5~\mathrm{\mu m}$ by the diffraction limited laser spot size, could also be improved to $\sim10~\mathrm{nm}$ with super resolution techniques \cite{Khatri2021}, or by using Fourier magnetic imaging\cite{Arai2015}.

Overall, this technique presents a competitive solution to the problem of sensing GHz magnetic fields, and with further development could utilise the two-dimensionality of the host material to dramatically enhance sensitivity, or even gain access to otherwise undetectable signals. This capability could open the door to sensing across diverse application areas, from utilising the materials low cytotoxicity \cite{Merlo2018} to probe biomolecular dynamics \cite{Zhang2021_PhysRevB, Shi2015} to collective excitations in magnetic thin films for novel computing architectures \cite{Bertelli2020, Prananto2021}.

\backmatter

\bmhead{Acknowledgments}

This work was supported by the Engineering and Physical Sciences Research Council [Grant numbers EP/S001557/1 and EP/L015331/1], Partnership Resource Funding from the Quantum Computing and Simulation Hub [EP/T001062/1] and an Engineering and Physical Sciences Research Council iCASE in partnership with Oxford Instruments Plasma Technology. Ion implantation was performed by Keith Heasman and Julian Fletcher at the University of Surrey Ion Beam Centre.



\section{{Methods}}
The sample consists of a chromium/gold (5/170 $\mathrm{nm}$ thick) coplanar waveguide (CPW), with a $10 ~\mathrm{\mu m}$ wide central conductor, on a sapphire substrate and designed to be matched to $50~\mathrm{\Omega}$ at 4 GHz. An hBN flake, approximately 100 $\mathrm{nm}$ thick, is placed on top of the CPW using the PDMS transfer method. Boron vacancies are generated/activated using C ion irradiation with an energy of 10 keV and dose of $1\times 10^{14}~\mathrm{cm^{-2}}$. Further details can be found in Baber \textit{et al.}\cite{NanoLett2022_Baber}.

PL is excited using a 488 $\mathrm{nm}$ diode laser, modulated by an acousto-optic modulator. The laser is coupled to a long working distance objective lens (N.A.=0.8) which focuses the light to a diffraction-limited spot $\sim0.5~\mathrm{\mu m}$ in diameter. The wavelength is selected to be close to the maximum absorption efficiency \cite{NanoLett2022_Baber}. Photoluminescence is collected with the same objective, separated from the excitation laser with a dichroic beamsplitter and further filtered by a 750 $\mathrm{nm}$ long pass filter. A fiber-splitter couples the light to a pair of single photon avalanche diodes (SPAD), which effectively doubles the count rate before saturation. The PL intensity is recorded using a time-correlated single photon counting module (Swabian Time Tagger 20), which has a data transfer limit of 8.5 Mtags/s that ultimately limits the signal-to-noise ratio that is achieved. The microwave control waveforms are generated using an arbitrary waveform generator (Keysight M8195A) and amplifier (30 $\mathrm{dB}$ amplification, maximum output power 30 $\mathrm{dBm}$) and are applied via a circulator to one end of the CPW. The signal source (Agilent 8648C) is connected to the other end of the CPW, via a second circulator. There is no clock synchronization between the control and signal sources. The optical and microwave excitation, APD gating and photon collection are synchronised using a digital pattern generator (Swabian Pulse Streamer). All measurements are performed with a DC magnetic field along the z-axis of $\sim40~\mathrm{mT}$ generated by a permanent magnet.

The results presented are from two similar devices. Device A is used to collect the data in Figures \ref{fig2} to \ref{fig4}, with device B used to collect the magnetic field mapping data in Fig \ref{fig5}. Due to a small difference in the sample heights relative to the permanent magnet, $\omega_0=(2\pi)2.52~\mathrm{GHz}$ for device A and $\omega_0=(2\pi)2.55~\mathrm{GHz}$ for device B.

\newpage
\section{Supplementary Information}
\subsection{\label{sec:13}{Sensor Calibration}}

\begin{figure*}[ht] 
\includegraphics[width=0.5\textwidth]{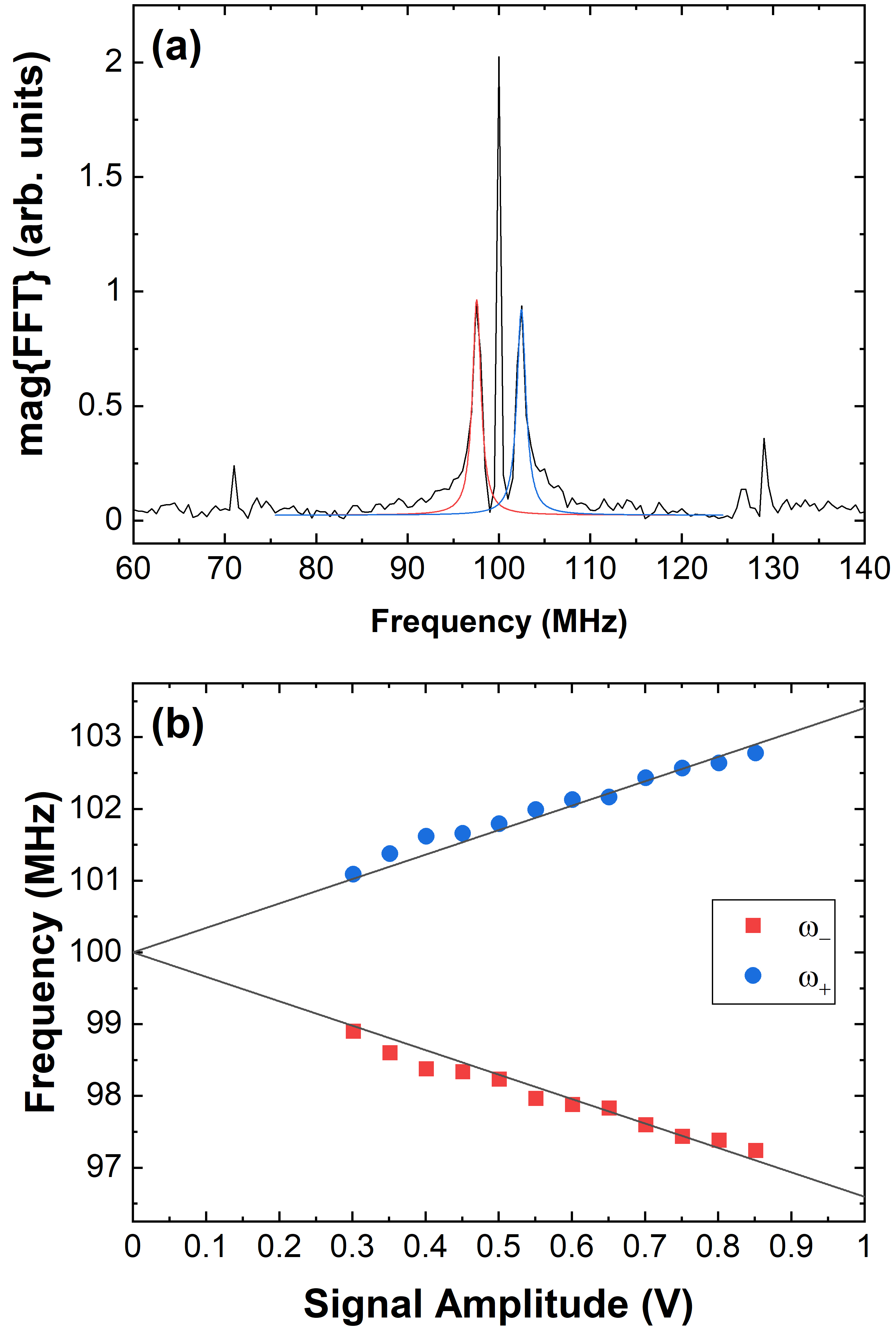}
\centering
\caption{\label{fig_SI_SigCal} (a) FFT spectrum of a CCDD Rabi-oscillation in the presence of a signal with amplitude 700 mV. (b) Frequency of signal induced sidebands as a function of the signal amplitude. The solid gray lines show linear fits to the data from which the signal amplitude is calibrated.}
\end{figure*}

To quantify the sensitivity it is necessary to calibrate the signal generator used to apply the signal. This is done through a series of experiments similar to those shown in Fig. \ref{fig2}. The CCDD Rabi-oscillation is measured for different in-plane signal amplitudes, $A$, applied to the CPW. From the FFT of this measurement the frequency of the Mollow-triplet sidebands ($\omega_+$ and $\omega_-$) are extracted using a Lorentzian fit (Fig. \ref{fig_SI_SigCal}(a)) and are plotted as a function of signal amplitude in Fig. \ref{fig_SI_SigCal}(b), showing the expected linear behaviour ($g/2=\omega_+-\omega_-\propto A$). This allows the signal amplitude to be converted from Volts to Tesla. A similar procedure is used to calibrate the response to out-of-plane signals and for the magnetic field mapping experiments when the signal is delivered via the external wire loop. 

\subsection{\label{sec:10}{Rotating Frame Transformations of an AC Signal}}

In the presence of an AC signal, the Hamiltonian gains an additional term, $H_{AC}$. The sensor response is different for in-plane and out-of-plane magnetic fields and so we treat the two cases separately:

\begin{eqnarray}
H_{AC\perp}= g_{\perp} \sigma_x \cos(\omega_st + \phi_s) \nonumber \\ 
H_{AC\parallel} = g_{\parallel} \sigma_z \cos(\omega_st + \phi_s)\nonumber
\end{eqnarray}

To define the signal field in the protected frame, we apply the same rotating frame transformations with respect to the control field drive frequencies. Starting with the electron spin resonance along the x axis, $\sigma_x \omega$, we find $H_c^{\prime}=e^{i\omega t\sigma_z/2}H_ce^{-i\omega t\sigma_z/2}$.

\begin{eqnarray}
H_{AC\perp}^{\prime}= \frac{g_x}{2}(\sigma_x^{\prime}\cos((\omega_s - \omega)t + \phi_s) + \sigma_y^{\prime}\sin((\omega_s - \omega)t + \phi_s)) \nonumber \\ 
H_{AC\parallel}^{\prime}=g_z\sigma_z \cos(\omega_s + \phi_s) \nonumber
\end{eqnarray}

where $^{\prime}$ denotes the first rotating frame and we have applied the RWA. $H_{AC\parallel}$ commutes with the transformation and so is unaffected. This produces a difference of $\omega$ in the resonant frequencies of the in-plane and out-of-plane signal fields. Applying the second transformation with respect to $\omega_m \sigma_x^{\prime}$, we find $H_c^{\prime\prime}=e^{i\omega_m t\sigma_x^{\prime}/2}H_ce^{-i\omega_m t\sigma_x^{\prime}/2}$.

\begin{multline*}
H_{AC\perp}^{\prime\prime} = \frac{g_x}{2}(\sigma_x^{\prime\prime}\cos((\omega_s - \omega)t + \phi_s) + \sigma_y^{\prime\prime}(\sin((\omega_s - \omega + \omega_m)t + \phi_s) + \sin((\omega_s - \omega - \omega_m)t + \phi_s)) - ...\\ \nonumber
...\sigma_z^{\prime\prime}(\cos((\omega_s - \omega - \omega_m)t + \phi_s) - \cos((\omega_s - \omega + \omega_m)t + \phi_s)))
\end{multline*}
\begin{multline*}
H_{AC\parallel}^{\prime\prime} = \frac{g_z}{2}(\sigma_z^{\prime\prime}(\cos((\omega_s - \omega_m)t + \phi_s) + \cos((\omega_s + \omega_m)t + \phi_s)) + ...\\ ...\sigma_y^{\prime\prime}(\sin((\omega_s + \omega_m)t + \phi_s) - \sin((\omega_s - \omega_m)t + \phi_s)))
\end{multline*}

where $^{\prime\prime}$ denotes the protected frame. To arrive at a more concise form, we apply a counter clockwise rotation around the $y^{\prime\prime}$ axis for $H_{AC\perp}^{\prime\prime}$ ($\sigma_x^{\prime\prime} \rightarrow \sigma_z^{\prime\prime}, -\sigma_z^{\prime\prime} \rightarrow \sigma_x^{\prime\prime}, \sigma_y^{\prime\prime} \rightarrow \sigma_y^{\prime\prime}$) and a clockwise rotation around the $y^{\prime\prime}$ axis for $H_{AC\parallel}^{\prime\prime}$ ($\sigma_z^{\prime\prime} \rightarrow \sigma_x^{\prime\prime}, \sigma_y^{\prime\prime} \rightarrow \sigma_y^{\prime\prime}$).

\begin{multline*}
H_{AC\perp}^{\prime\prime} = \frac{g_x}{2}(\sigma_z^{\prime\prime}\cos((\omega_s - \omega)t + \phi_s)) + \frac{g_x}{4}(S_+e^{-i((\omega_s - \omega - \omega_m)t + \phi_s)} - S_-e^{-i((\omega_s - \omega + \omega_m)t + \phi_s)} + h.c.) \\
H_{AC\parallel}^{\prime\prime} = \frac{g_z}{2}(S_+e^{-i((\omega_s - \omega_m)t + \phi_s)} + S_-e^{-i((\omega_s + \omega_m)t + \phi_s)} + h.c.)
\end{multline*}

For the sensor to detect the signal it must be resonant with the energy gap $\epsilon_m$ in the protected frame. There are eight possible ways of achieving this using these five resonance conditions (see \ref{sec:level4} for more details). A ninth resonance, observed for signal frequencies of $\omega_s = \epsilon_m$, appears after a third rotating frame transformation.

\subsection{\label{sec:12}Additional sensor resonance at the drive frequency}

An additional resonance is observed in Fig. \ref{fig4}a when $\omega_s = \epsilon_m$. This condition does not appear in Eq. \ref{eq:ProtectedPerpFrameSignal} or \ref{eq:ProtectedParallelFrameSignal}, but can be observed after an additional frame transformation. Combining the protected frame control and out-of-plane signal Hamiltonians, $H^{\prime\prime} = H_C^{\prime\prime} + H_{AC}^{\prime\prime}$, we find:

\begin{multline*}
H^{\prime\prime} = \frac{1}{2}\epsilon_m \sigma_z^{\prime\prime} + \frac{g_z}{2}[\sigma_z^{\prime\prime}(\cos((\omega_s - \omega_m)t + \phi_s) + \cos((\omega_s + \omega_m)t + \phi_s)) + \nonumber \\
\sigma_y^{\prime\prime}(\sin((\omega_s + \omega_m)t + \phi_s) - \sin((\omega_s - \omega_m)t + \phi_s))]
\end{multline*}

where we have chosen $\theta_m = 0$ and the resonant case $\Omega = \omega_m$. Moving to a third rotating frame at an arbitrary frequency f, along $\sigma_z^{\prime\prime}$ we arrive at $H^{\prime\prime\prime}=e^{ift\sigma_z^{\prime\prime}/2}H^{\prime\prime}e^{-ift\sigma_z^{\prime\prime}/2}$

\begin{multline*}
H^{\prime\prime\prime} = \frac{1}{2}(\epsilon_m - f)\sigma_z^{\prime\prime\prime} + \frac{g_z}{2}(\sigma_z^{\prime\prime}\cos((\omega_s - \omega_m)t + \phi_s) + \cos((\omega_s + \omega_m)t + \phi_s)) + \\
\frac{1}{2}\sigma_y^{\prime\prime\prime}(\sin((\omega_s + \omega_m + f)t + \phi_s) + \sin((\omega_s + \omega_m - f)t + \phi_s) + \cos((\omega_s + \omega_m - f)t + \phi_s) - \cos((\omega_s + \omega_m + f)t + \phi_s) - \\
\sin((\omega_s - \omega_m + f)t + \phi_s) - \sin((\omega_s - \omega_m - f)t + \phi_s) - \cos((\omega_s - \omega_m - f)t + \phi_s) + \cos((\omega_s + \omega_m + f)t + \phi_s)))
\end{multline*}

Choosing the condition $f = \omega_m/2$ yields the signal terms, $\frac{1}{4}g_z\sigma_y^{\prime\prime\prime}(\sin((\omega_s - \frac{1}{2}\omega_m)t + \phi_s) - \cos((\omega_s - \frac{1}{2}\omega_m)t + \phi_s)$. When $\omega_s = \epsilon_m$, these terms are resonant with the eigenenergies of the frame, $\frac{1}{2}\sigma_z^{\prime\prime\prime}(\epsilon_m - \frac{1}{2}\omega_m)$, and therefore drive a transition between spin states, producing the resonance observed in Fig. \ref{fig4}(a).

\subsection{\label{sec:11}Biot-Savart Law for Two Parallel Conductors}

The plot in Figure \ref{fig5}(c) uses an analytical solution to the Biot-Savart law for two parallel conductors, which we include below:

\begin{eqnarray}
B_\perp = \frac{\mu_0}{2\pi} ((\frac{x-x_0}{(x-x_0)^2 + z_0^2} - \frac{x-x_1}{(x-x_1)^2 + z_1^2})I_\perp\hat{i} \nonumber
\label{eq:BiotSavartx}
\end{eqnarray}
\begin{eqnarray}
B_\parallel = \frac{\mu_0}{2\pi} ((\frac{z}{(x-x_0)^2 + z_0^2}-\frac{z}{(x-x_1)^2 + z_1^2})I_\parallel\hat{k} \nonumber
\label{eq:BiotSavartz}
\end{eqnarray}

where $\mu_0$ is the vacuum permeability, $I$ is the current, $x$ and $z$ denote the positions of the sensor relative to the wire, $x_0, x_1, z_0$ and $z_1$ denote the absolute positions of the wire.






\end{document}